\documentclass{emulateapj}
\newcommand{\etal}{et al.\ }
\newcommand{\etalb}{et al.}
\newcommand{\be}{\begin{equation}}
\newcommand{\ba}{\begin{eqnarray}}
\newcommand{\ee}{\end{equation}}
\newcommand{\ea}{\end{eqnarray}}

\begin{document}
\title{The imprint of dissipation on the shapes of merger remnant LOSVDs}

\author{Loren Hoffman\altaffilmark{1}, Thomas
J. Cox\altaffilmark{2}, Suvendra Dutta\altaffilmark{2},
Lars Hernquist\altaffilmark{2}}

\email{l-hoffman@northwestern.edu}

\altaffiltext{1}{Department of Physics and Astronomy, Northwestern 
University, Dearborn Observatory, 2131 Tech Drive, Evanston, IL, 60208}

\altaffiltext{2}{Department of Astronomy, Harvard University, 60 
Garden Street, Cambridge, MA 02138}

\begin{abstract} 
The properties of elliptical galaxies are broadly consistent with 
simulated remnants of gas-rich mergers between spirals, motivating more 
detailed studies of the imprint of this formation mechanism on the remnant 
distribution function.  Gas has a strong impact on the non-Gaussian shapes
of the line-of-sight velocity distributions (LOSVDs) of the merger remnant,
owing to the embedded disk that forms out of the gas that retains its angular
momentum during the merger, and the strong central mass concentration from 
the gas that falls to the center.  The deviations from Gaussianity are 
effectively parametrized by the Gauss-Hermite moments $h_{3}$ and $h_{4}$, 
which are related to the skewness and kurtosis of the LOSVDs.  We quantify the 
dependence of the $(h_{3},h_{4})-v / \sigma$ relations on the initial gas 
fraction $f_{gas}$ of the progenitor disks in 1:1 mergers, using Gadget-2 
simulations including star formation, radiative cooling, and feedback 
from supernovae and AGN.  For $f_{gas} \lesssim 15$\% the overall 
correlation between $h_{3}$ and $v / \sigma$ is weak, consisting of a 
flat negatively correlated component arising from edge-on viewing angles 
plus a steep positively correlated part from more face-on projections.
The spread in $v / \sigma$ values decreases toward high positive 
$h_{4}$, and there is a trend toward lower $h_{4}$ values as $f_{gas}$ 
increases from 0 to 15\%.  For $f_{gas} \gtrsim 20$\% the 
$(h_{3},h_{4})-v / \sigma$ distributions look quite different - there is a 
tight negative $h_{3}-v / \sigma$ correlation, and a wide spread in 
$v / \sigma$ values at all $h_{4}$, in much better agreement with 
observations.  Re-mergers of the high-$f_{gas}$ remnants (representing dry 
mergers) produce slowly rotating systems with near-Gaussian LOSVDs.  We 
explain all of these trends in terms of the underlying orbit structure of the 
remnants, as molded by their dissipative formation histories.
\end{abstract}


\section{Introduction}

Many of the observed properties of elliptical galaxies indicate a violent 
formation history.  They are dynamically hot systems, with high velocity 
dispersions dominating over ordered stellar streaming.  Gas-rich tidal tails,
and rings and shells indicative of the recent disruption of a spiral 
galaxy, often surround systems otherwise resembling ordinary giant 
ellipticals \citep{a66,vd05}.  These observations first led \citet{tt72} to 
suggest that elliptical galaxies are the products of mergers between spirals.
N-body simulations of mergers between disk galaxies generically 
reproduce many of the gross properties of observed ellipticals, including 
density profiles that follow the ``$r^{1/4}$'' law in projection 
\citep{va82,mg84}, flat rotation curves \citep{w78,fs82}, slow rotation and 
anisotropic velocity distributions \citep{w78,g81}, and triaxial shapes 
\citep{g81,b88}.  Today we understand the Toomre merger hypothesis within the 
broader context of hierarchical structure formation in the Universe - in 
the favored $\Lambda$CDM cosmology, an elliptical galaxy-sized halo has 
typically suffered $\sim$1 major merger since $z \sim 2-3$, the epoch over
which the bulk of its stellar mass formed (e.g. 
\citealt{lc93,ms96,k96,d06,hop08a,hop09a,s08}).

Observations of elliptical galaxies also show signatures of dissipation in 
their buildup.  The phase space densities in the cores of ellipticals exceed 
the maximum densities in spirals, implying a violation of Liouville's theorem 
if ellipticals arise from {\it dissipationless} spiral-spiral mergers 
\citep{c86}.  This problem is naturally resolved by a gas component that is 
driven to the center by tidal torques during the merger 
\citep{nw83,bh91gas,hsh93}.
Elliptical galaxies also display kinematic features that are most easily 
explained as signatures of dissipation, such as rapid oblate rotation 
(e.g. \citealt{d83,b92}), embedded disks \citep{b90,rw90}, and kinematically 
decoupled subsystems \citep{hb91kdc,e82,fi88,js88,sauron12}.  \citet{tjkin} 
showed that a wide variety of the photometric and kinematic properties of 
observed ellipticals, including their half-light radii and velocity 
dispersions, flattening, isophotal shapes, velocity anisotropy, and major and 
minor axis rotation were much better reproduced by an ensemble of simulated 
40\% gas disk merger remnants than by the corresponding ensemble of 
dissipationless mergers.  \citet{r06fp} found that a gas fraction of 
$\sim$30\% was needed to match the observed tilt in the ``fundamental plane'' 
\citep{d87,dd87}, a value close to the typical gas fraction at $z\sim2-3$.  

The violent relaxation in galaxy mergers is {\it incomplete} - substantial 
memory of the initial conditions is retained.  For example the 
remnant may tend to rotate in the sense of the initial orbital angular 
momentum or disk spins \citep{barnes92,hernquist92,hernquist93},
and an initial stellar metallicity 
gradient will be blurred but not erased during the merger \citep{w78}.  This 
incompleteness is enhanced by the presence of gas, which can form cold 
features in the stellar distribution relatively late in the merger process, 
which subsequently experience less violent relaxation than features present 
at the beginning of the merger.  This fine structure provides a fossil 
record of the galaxy's formation history (e.g. \citealt{statler}).

One way to parametrize the finer features in a galaxy's distribution 
function, which is well-suited to spectroscopic observations, is through the 
moments of the line-of-sight velocity distribution (LOSVD) as a function of
projected location in the galaxy.  For continuous stellar line profiles, 
these moments are measured by fitting the LOSVD to a Gaussian multiplied by
a truncated series of Gauss-Hermite (GH) basis functions $H_{k}(w)$, 
\begin{equation}
F_{GH} (v_{los}) \propto e^{-\frac{1}{2} w^{2}} [1+\sum_{k=3}^{n}h_{k}H_{k}(w)],
\end{equation}
where $w \equiv (v_{los}-\bar{v})/\sigma$ and $\sigma$ is the line-of-sight 
(LOS) velocity dispersion.  This method is less sensitive to noise than 
computing the 
moments directly \citep{vf93,g93,BandM}.  For modest deviations from 
Gaussianity, $h_{3}$ represents the skewness of the distribution (see 
\citealt{BandM}, Fig. 11.5) and can, for example, indicate whether the net
rotation in a galaxy arises from the dominant orbit population, or a small
group of streaming stars.  The
quantity $h_{4}$ measures the kurtosis; positive 
$h_{4}$ values generally indicate a radial anisotropy in the velocity 
distribution, while negative $h_{4}$ indicates a tangential bias \citep{g93}.
In a non-rotating system, $h_{4}$ can distinguish between a dominant 
population of radial (box) orbits, and cancelling streams of high-angular 
momentum (tube) orbits.

In recent years, integral field spectroscopy (IFS) has made it possible to
obtain high S/N, high-resolution 2D maps of the LOSVD in nearby ellipticals
\citep{sauron1,sauron3}.  Schwarzschild modelling studies have
shown that, in practice, the observed 2D maps of the first four moments of
the velocity distribution, $\{v, \sigma, h_{3}, h_{4}\}$, are sufficient
to uniquely reconstruct the full 3D stellar orbital distribution in most
cases \citep{saurdynmod,schwarzmod,ngc4365,vv08}.  Patterns in the higher 
moments, $\{h_{k}, k \geq 5\}$, typically fall below the noise in the 
observations.  It is therefore standard practice to truncate the GH series at 
$k=4$, and encapsulate the non-Gaussian shape of the distribution in the two 
parameters $h_{3}$ and $h_{4}$.    

Previous authors \citep{bb00,jess05,naab06,jess07} have observed that 
simulated gas-rich merger remnants occupy different areas of the 
$h_{3,4}-v/\sigma$ planes from dissipationless disk-disk merger remnants.  
Dissipationless remnants tend to show an overall positive correlation 
between $h_{3}$ and $v/\sigma$, while in gas-rich remnants, as in observed 
ellipticals \citep{b94,sauron12}, this correlation is negative. 
\citet{gg06} show that adding a bulge component to the merging disks can also 
yield larger $h_{3}$ values and a negative $h_{3}-v/\sigma$ correlation, by 
allowing the disks to retain more of their initial angular momentum.  
\citet{b94} demonstrate that the observed $h_{3}-v / \sigma$ relation is too 
steep to be explained by two-integral oblate rotator models (e.g. 
\citealt{dg94}), but can easily be accounted for by the superposition of a 
hot stellar spheroid with a cold embedded disk.  This type of distribution 
function arises naturally from a combination of violent relaxation and 
dissipation in gas-rich mergers of spiral galaxies \citep{h09b}.

\citet{jess05} and \citet{naab06} performed a detailed study of 
the stellar orbit structure of 1:1 and 3:1 merger remnants with 0 and 10\% 
gas, and its imprint on their photometric and kinematic properties.  They 
found that the gas drives an exchange between box and short-axis tube orbits, 
making the remnants more oblate in shape.  This suppression of box orbits 
strongly influences the shapes of the LOSVDs, bringing the $h_{3,4}-v/\sigma$ 
relations into better agreement with observed rapidly rotating ellipticals.  
\citet{jess07} showed that 2D kinematic maps of the same set of remnants
display many of the intriguing features seen in real galaxies, including 
misaligned rotation, embedded disks, and kinematically decoupled cores.

In this paper we explore the effect of gas on the $h_{3,4}-v/\sigma$
relations using a version of the tree/SPH code Gadget-2 \citep{gadget2} that 
includes star formation, radiative cooling, and feedback from supernovae and 
AGN \citep{gadsf, gadfeedbak}.  The inclusion of real-time star formation
allows us to consider the higher gas fractions typical of spiral galaxies at 
the peak elliptical formation epoch, $f_{gas} \sim 30\%$ at $z \sim 2$.  For
fixed $f_{gas}$, the dissipative signature on the dynamics may be reduced 
by star formation, if the gas is converted to collisionless material early on 
in the merger.  The dissipational features may also be more spatially 
extended when star formation is included.

We simulate a representative series of mergers between equal-mass, Milky 
Way-sized disk galaxies at a series of initial gas fractions ranging from 0 
to 40\%, as well as re-mergers of the spheroidal remnants.  We quantify the 
dependence of the $h_{3,4}-v/\sigma$ diagrams on $f_{gas}$, and explain the 
patterns that we find in terms of the underlying stellar orbital 
distribution.  Our primary goals are: (a) to see whether this diagnostic 
points to the same typical gas fraction as other indicators, such as the FP 
tilt and peak quasar redshift; (b) to locate ``wet'' (gas-rich 
disk-disk) and ``dry'' (re-mergers of gas-poor spheroid) mergers in 
$(h_{3,4}-v/\sigma)$ space, with the aim of distinguishing these two 
populations in IFS observations; and (c) to connect the observed trends in 
the non-Gaussian moments with the underlying orbit structure typical of 
merger remnants in an intuitive way.
 
\section{Simulations and methods}

Our initial galaxy models consist of exponential disks embedded in 
Hernquist (1990) dark 
matter halos, with masses comparable to the Milky Way.  The galaxy
models are described in detail in \citet{gadfeedbak}.  The halo concentration
and spin parameter were set to $c=0.9$ and $\lambda=0.033$.  The disk mass was 
4.1\% of the halo mass, and its specific angular momentum was assumed equal 
to that of the halo.  The halo was realized with 120,000 particles, and 
80,000 stars initially comprised the disk.  A fraction $f_{gas}$ of the 
disk stars were replaced with SPH gas particles at the start of the 
simulation, and this component could subsequently form new stars if a 
threshold density $\rho_{th}$, tuned to match the Schmidt Law 
\citep{schmidt,kenn}, was exceeded.  The gravitational softening 
length was 140 pc, which set a minimum spatial resolution scale in the core.
(See also \citealt{tjkin} for further description of the simulations.)

The disks began at a separation of 100kpc, on parabolic orbits with a 
periapsis distance of 7.1kpc.  They typically merged after around 1.5Gyrs, and
the simulations were run to a final time $t_{end}=3h^{-1}=$4.3 Gyrs.  
\citet{tjkin} show that the global remnant properties reach a steady state at 
one effective radius ($R_{e}$) around 0.3Gyrs after the merger, so by 
$t_{end}$ the remnants can safely be considered relaxed at this scale.  The 
set of eight merger trajectories used in this study correspond to orbits 
$i-p$ in \citet{tjkin}, chosen to uniformly sample the phase space of 
possible orbits \citep{barnes92}.  

The series of eight representative disk-disk mergers was performed at seven 
different gas fractions, $f_{gas}=$0, 5, 10, 15, 20, 30, and 40\%.  We also
ran two series of re-mergers of the 20 and 40\% gas remnants,
intended to represent dry mergers of realistic gas-poor ellipticals.
To produce a representative sample for each re-merger series, we randomly
selected eight pairs from among the disk-disk merger remnants, and re-merged
them on trajectories $i-p$ again.

For each of the simulated remnants we selected 100 random isotropically 
sampled viewing angles, and a synthetic 2D map of the kinematics was 
constructed within $1R_{e}$ for each LOS.  A constant mass-to-light ratio 
was assumed throughout the analysis.  In each spatial bin we constructed a 
velocity histogram and extracted $v$, $\sigma$, $h_{3}$, and $h_{4}$ by 
performing a least-squares fit to the 5-parameter function
\begin{eqnarray*}
F(y) = A \exp [-\frac{(y-v)^{2}}{2 \sigma^{2}} ] \{ 1 & + & h_{3} H_{3} (\frac{y-v}{\sigma}) \\
& + & h_{4} H_{4} (\frac{y-v}{\sigma}) \}
\end{eqnarray*}
using the Levenberg-Marquardt method \citep{numrec}.  We used 40 $\times$ 40
spatial bins and 80 velocity bins within $\pm 3.5 v_{rms}$.  The Gadget 
particles were smoothed over a radius $h_{smooth}=\max(h_{see},h_{ngb})$, 
where $h_{see} = 150 pc$ corresponds to a seeing of 1.5" at 20Mpc, and 
$h_{ngb}$ is a smoothing length based on the distance to the 128th nearest 
neighbor.  This ensured that the effective particle count in each spatial bin 
was $\gtrsim$ 1000.  At $t_{end}$ we also froze the potential, expanded it in 
the basis set of \citet{scf}, and classified the orbits of the stellar 
particles in this potential.  The results of this exercise are thoroughly 
presented in \citep{myorb}.

\section{Results}

\subsection{The $h_{3}-v / \sigma$ distribution}

\begin{figure*}
\epsscale{1.17}
\plotone{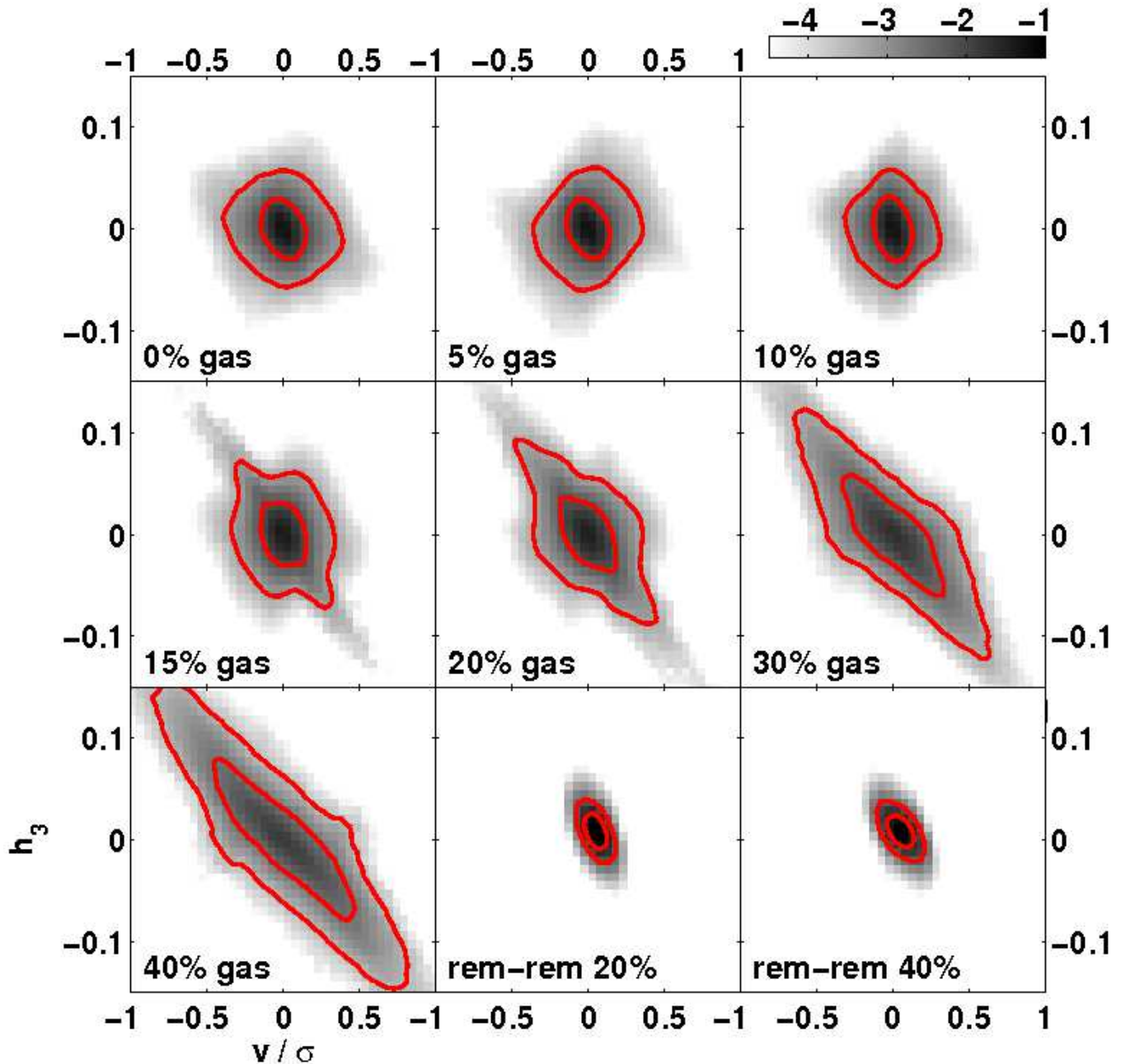}
\caption{Local $h_{3}-v/\sigma$ relation for each of our nine merger simulation sets.  The first seven panels are for disk-disk mergers of different gas fractions, while the last two panels (lower right) are for dry mergers of the 20 and 40\% gas remnants.  Each map is averaged over the eight unbiased merger orbits, and 100 isotropically selected lines of sight per remnant.  The shading is proportional to the logarithm of the luminosity in each $h_{3}-v/\sigma$ bin, normalized so that the entire map sums to one.  The red lines are contours containing 68 and 95\% of the luminosity.}
\vspace{0.1in}
\end{figure*}

Our results for the $h_{3}-v / \sigma$ relation are presented in Fig. 1.  The 
shape of the distribution is quite sensitive to $f_{gas}$ for the disk-disk 
merger remnants.  At low gas fractions the overall correlation between $h_{3}$
and $v / \sigma$ is weak; the distribution has a diamond shape 
consisting of a flat, negatively correlated component plus a steep, positively
correlated one.  Although the intrinsic structure changes rapidly from 
flattened, prolate-triaxial systems dominated by box orbits to rounder, more
oblate systems dominated by short axis ($z-$) tube orbits as the gas fraction 
increases from 0 to 15\% \citep{myorb}, there is no strong signature of this 
transformation in the asymmetric $h_{3}$ moment.

For $f_{gas} \gtrsim 20\%$ the shape of the $h_{3}-v / \sigma$ distribution 
changes rather abruptly.  These gas-rich remnants are characterized by high 
$h_{3}$ moments that are strongly anti-correlated with $v / \sigma$, in better
agreement with observed fast-rotator ellipticals \citep{b94,sauron12}.  The 
observations thus favor typical progenitor gas
fractions $ \gtrsim$ 25\%, in agreement with other indicators such as the
shapes of the remnants \citep{tjkin,myorb}, the two-component structure of
ongoing mergers \citep{hop08b}, cuspy \citep{hop09c} and cored
ellipticals \citep{hop09d}, and the tilt of the FP 
\citep{r06fp,hop08c}.  In the SAURON data there is also evidence for a second 
population of fast rotators with low $h_{3}$ values \citep{sauron12};
these systems cannot be explained by the simulations performed in this study.

Re-mergers of the high-gas fraction remnants still show a distinct negative 
correlation between $h_{3}$ and $v / \sigma$, but the remnant LOSVDs are much
more concentrated near the origin of the $h_{3}-v / \sigma$ plane.  Note the
much lower spread in re-merger $h_{3}$ values at $v=0$ compared to disk-disk
mergers of any gas fraction.  This result is not surprising, since violent 
relaxation in the re-merger tends to blur out structure in the distribution 
function and drive the LOSVDs closer to a thermal distribution.  Fig. 1 
quantifies the degree of thermalization in a single dry merger.

\begin{figure*}
\epsscale{1.17}
\plotone{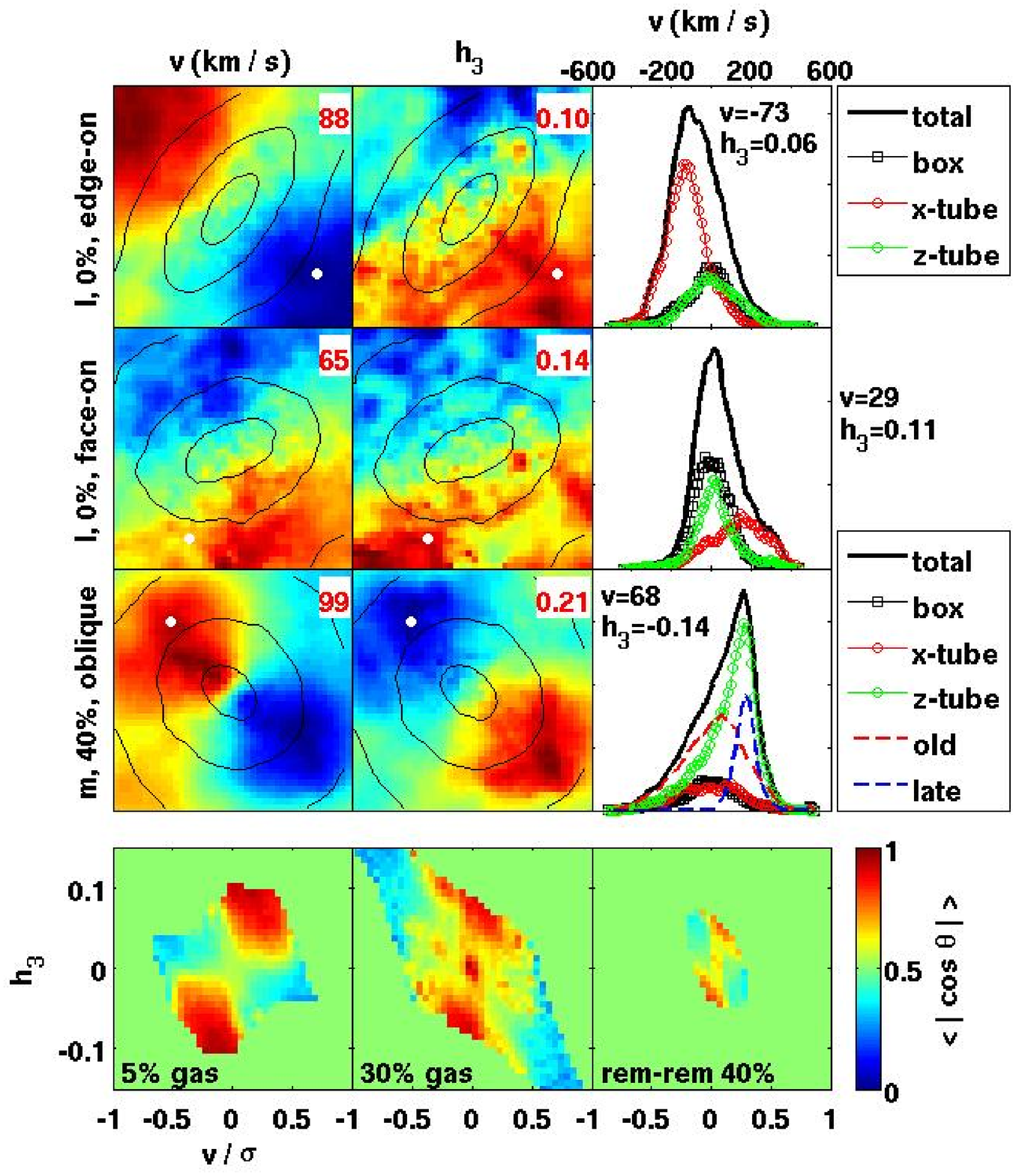}
\caption{Explaining the $h_{3}-v/\sigma$ trends.  {\it Top row:}  Merger orbit $l$, 0\% gas, at a nearly edge-on viewing angle.  {\it Second row:}  Same remnant, viewed nearly face-on.  {\it Third row:}  Merger orbit $m$, 40\% gas, viewed at an angle about 30$^{\circ}$ from edge-on.  {\it First two columns:} $v$ and $h_{3}$ maps within $1R_{e}$.  The color scale runs from $-h_{max}$ to $h_{max}$, where $h_{max}$ is the maximum value of the moment in any pixel, and the number in the upper right corner gives the value of $h_{max}$.  The black contours are over-plotted isophotes.  The third column shows the smoothed LOSVD at the location specified by the white dot on the $v$ and $h_{3}$ maps.  The heavy black line is the full LOSVD, while the red and green circles and black squares break it down by orbit class.  In the case of the gas-rich remnant (bottom row), we also show the separate contributions from old stars (red dashed line) and the 15\% of the new stars that formed latest (blue dashed line).  (See the text for further discussion of these plots.)  {Bottom row:}  Net contribution of face-on and edge-on projections to the $h_{3}-v/\sigma$ distributions.  The three maps correspond to the second, sixth, and ninth panels of Fig. 1, but now each bin is colored according to the luminosity-weighted mean $|\cos \theta|$, where $\theta$ is the polar angle between the LOS and the intrinsic short axis of the triaxial remnant.  $|\cos \theta| = 1$ means perfectly face-on, while $|\cos \theta| = 0$ is perfectly edge-on.}
\end{figure*}

In Fig. 2 we illustrate how the positive and negatively correlated components
arise in the low-$f_{gas}$ remnants, and how the strong $h_{3}-v / \sigma$ 
anti-correlation arises in the gas-rich ones.  In dissipationless remnant $l$
(top six panels), the major-axis ($x-$) tube orbits produce substantial minor 
axis rotation even though they only comprise $\sim$20\% of the stellar mass 
within 1$R_{e}$, since they are highly streaming.  In a pixel near the minor 
axis in an edge-on projection (looking along the intermediate axis), 
$x-$tubes dominate the luminosity, and when combined with the $z-$tube
and box orbits peaked at $v=0$ they result in a distribution with an extended
tail in the direction opposite the mean velocity.  Since the velocity 
distribution of the $x$-tube orbits is not too strongly peaked, the $h_{3}$ 
value is not very high and the $h_{3}-v / \sigma$ relation is relatively 
flat.  

When the same remnant is viewed face-on (projected along the short axis),  
the three orbit classes are all piled on top of each other, and the streaming 
$x-$tubes form a bump in the tail of the dominant distribution of boxes and 
$z-$tubes.  We thus get an extended tail in the same direction 
as the mean 
velocity, i.e. a positive $h_{3}-v/\sigma$ correlation.  This high-$v$ bump 
in the tail produces a stronger asymmetry in the LOSVD while the velocities 
in the face-on projection tend to be lower, so this effect produces a much 
steeper $h_{3}-v/\sigma$ relation than the effect illustrated in the top row.

The locus of face-on and edge-on projections in the $h_{3}-v / \sigma$ diagram,
averaged over all remnants of a given gas fraction, is shown in the bottom row
of Fig. 2.  To illustrate the trends we show one set of low-$f_{gas}$ 
disk-disk mergers, one set of high-$f_{gas}$ disk-disk mergers, and one set
of re-merger remnants.  As indicated in the above example, face-on 
projections dominate a steep positively-correlated component, while edge-on
projections dominate the shallower negatively correlated component.  The
positively correlated component is relatively suppressed at high gas 
fractions. 

In gas-rich remnants the dominant population ($z$-tube orbits) is streaming as
in the edge-on dissipationless remnant, only now the streaming is much stronger.
This effect arises from the embedded disk that re-forms after the merger out of
the gas that did not lose its angular momentum and fall to the center 
\citep{h09b}.  The red and blue dashed lines in the bottom LOSVD show the 
contribution from ``old'' stars put in as collisionless particles at the 
beginning of the simulation, and from the 15\% of the new stars (formed out of
gas during the merger) that formed latest.  These late-forming stars comprise
only a few percent of the total stellar mass, but are prominent in spatial bins
lying along the major axis.  These disk stars have a much lower velocity 
dispersion than the remnant as a whole, and produce a sharp high-$v$ peak
in the dominant $z-$tube population, giving a distribution that is strongly
skewed in the sense opposite the mean velocity.

As a rule, disk-disk mergers tend to produce large asymmetric deviations from
Gaussianity owing to subcomponents that retain memory of the initial streaming
of the disk stars, and to gas that is redistributed and spun up through 
dissipation to form a thin embedded disk. 
The $h_{3}-v / \sigma$ correlation is generally negative when the dominant 
orbit population tends to stream, and positive when the rotation is caused by
sub-dominant streaming population superimposed on the tail of the dominant 
orbits.

\subsection{The $h_{4}-v / \sigma$ relation}

\begin{figure*}
\epsscale{1.17}
\plotone{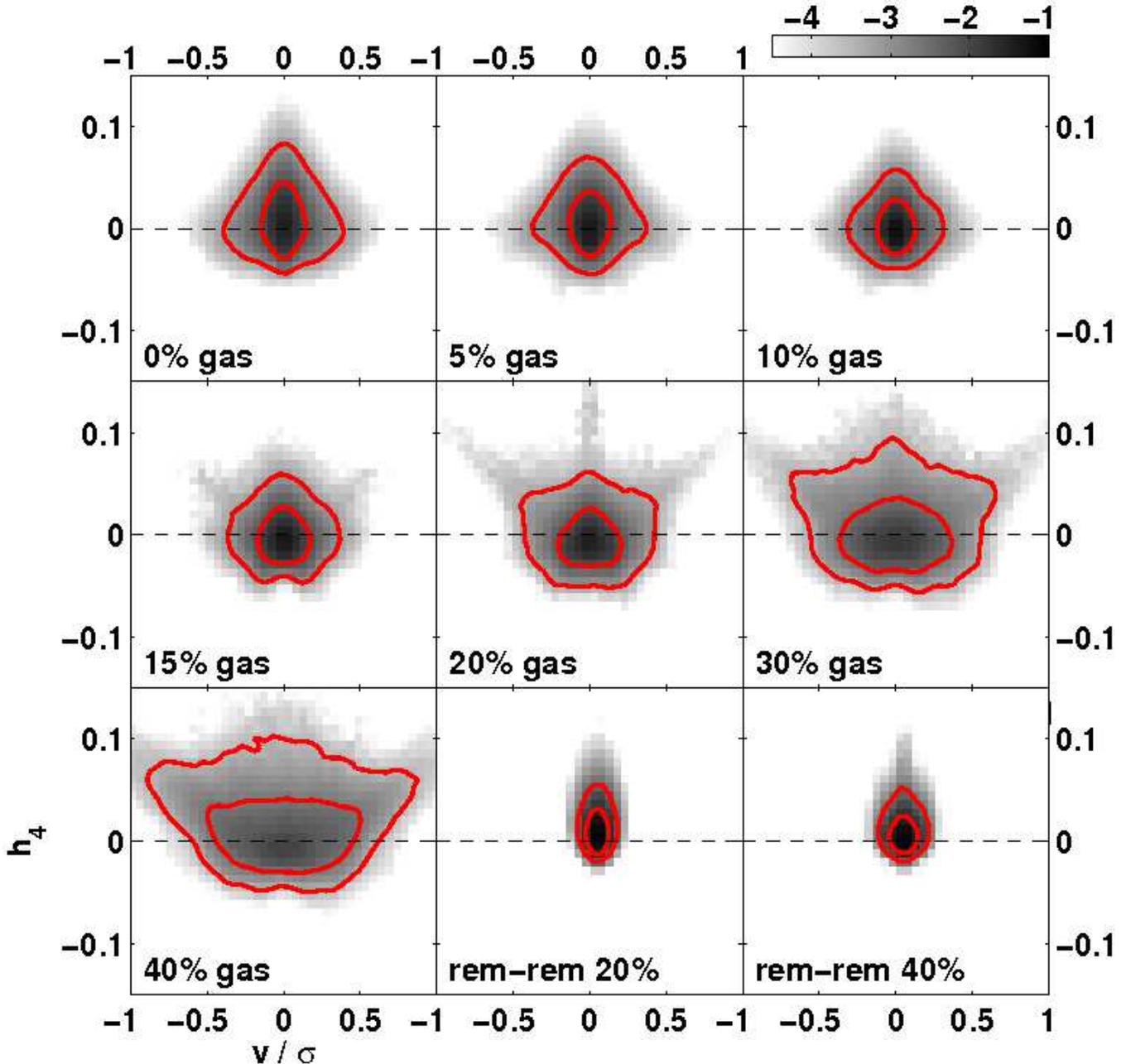}
\caption{Local $h_{4}-v/\sigma$ relation for each of our simulation sets.  The shading and contours are as in Fig. 1.}
\vspace{0.29in}
\end{figure*}

We now turn to the symmetric deviations of the LOSVDs from a Gaussian shape,
measured by the ``kurtosis'' parameter $h_{4}$.  In Fig. 3 we present the 
$h_{4}-v / \sigma$ relations from our simulations.  For 
$f_{gas} \lesssim 15$\% the $h_{4}-v / \sigma$ distribution has a triangular 
shape that narrows sharply toward high positive $h_{4}$.  Note that this is
the naive expectation for a single population of tube orbits - viewed edge 
on the high-angular momentum orbits tend to pile up in the tails of the 
velocity distribution and any significant imbalance in the sense of their 
rotation produces a high $v / \sigma$, while viewed face-on they pile up
in a peak around $v=0$.  However this pattern differs sharply from IFS 
observations, in which a mild increase in the $v / \sigma$ spread is actually 
seen toward high $h_{4}$ \citep{sauron12}.  There is also a trend toward 
lower $h_{4}$ values as $f_{gas}$ increases from 0 to 15\%, arising from the
conversion of box orbits into $z-$tubes by the central mass concentration 
formed by the gas that falls to the center and produces the starburst 
\citep{bh96,mh94,naab06}.

We get a marked change in the shape of the $h_{4}-v / \sigma$ distribution at 
$f_{gas}$ $\sim$ 20\%, as we did for the $h_{3}-v / \sigma$ relation.  At 
higher gas fractions we get strong positive as well as moderate negative 
$h_{4}$ values over a wide range in $v / \sigma$ - the $v / \sigma$ spread 
actually increases mildly toward high $h_{4}$ as in observed systems. 

\begin{figure*}
\epsscale{1.17}
\plotone{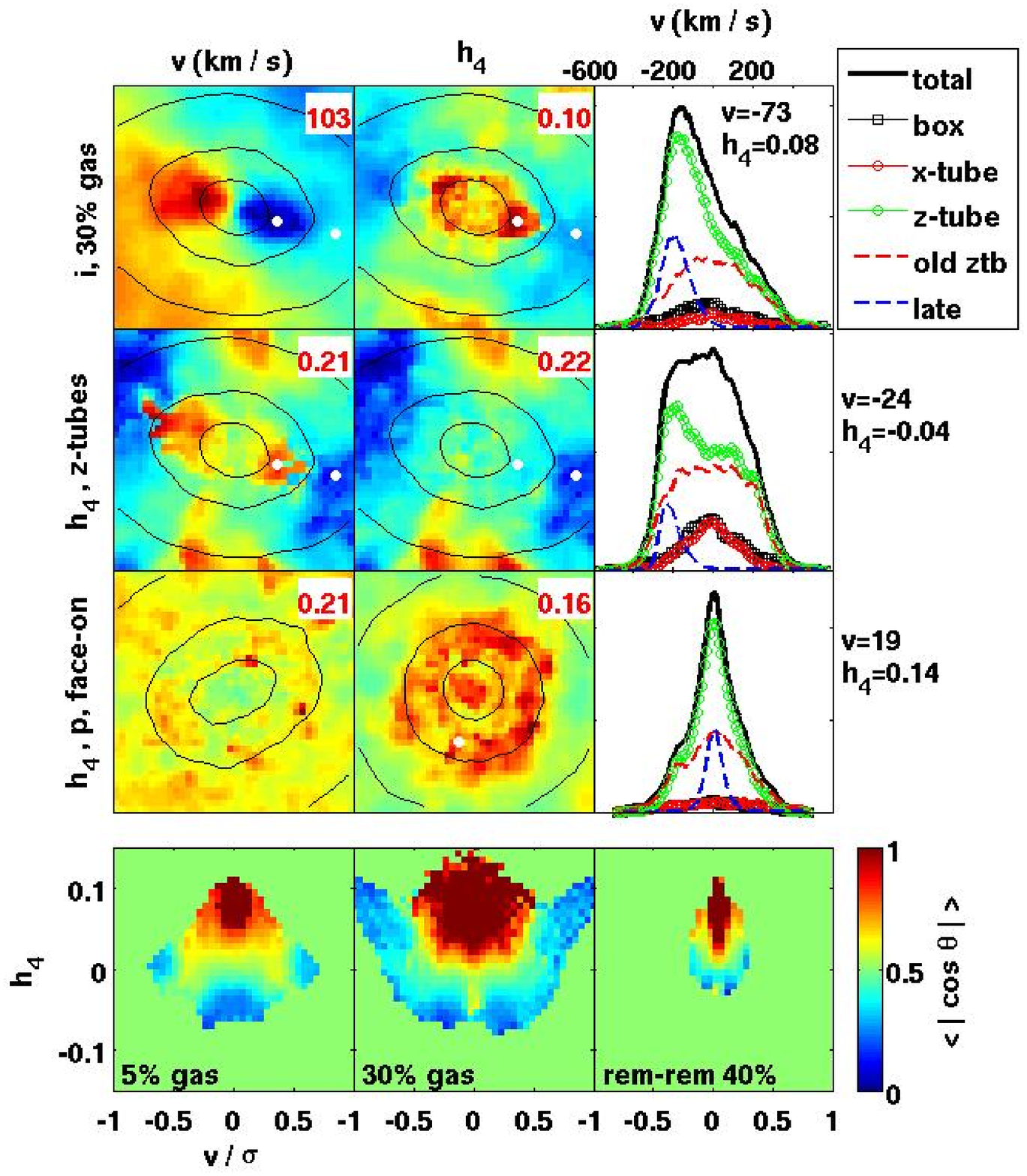}
\caption{Explaining the $h_{4}-v/\sigma$ relation.  The top six panels are from merger orbit $i$, viewed about 15$^{\circ}$ from edge-on.  {\it First row:} $v$ and $h_{4}$ maps for this remnant.  {\it Second row:} $h_{4}$ maps for the $z$-tube orbits alone.  The left map is for all of the stars on $z$-tube orbits, while the right map includes only {\it old} stars on $z$-tube orbits, excluding those that formed from gas during the merger.  In the right-hand column the LOSVDs are shown at the two points indicated by the white dots, broken down by orbit class and stellar age as in Fig. 2.  {\it Third row:} $h_{4}$ maps for merger orbit $p$, viewed face-on, at 0\% gas (left) and 40\% gas (right).  The LOSVD at the dotted location on the 40\% gas map is shown in the third column.  {\it Bottom row:} Net contribution of face-on and edge-on projections to the $h_{4}-v / \sigma$ distribution, plotted as in the bottom row of Fig. 2.}
\end{figure*}

Like the transformation in the $h_{3}-v / \sigma$ distribution, this rapid 
transition owes to the appearance of a strong disk component starting at
$f_{gas} \sim 20$\%, as shown in Fig. 4.  In high-$v/ \sigma$, edge-on 
projections of the disk, $h_{4}$ is influenced by two competing effects -
the strong negative kurtosis of the more isotropic population of old stars 
on $z-$tube orbits, and the strongly peaked population of late-forming disk
stars that tend to produce a positive $h_{4}$ in superposition with the old
component.  In the pixel shown in the top row of Fig. 4, the latter effect 
wins out.  Although the old stellar $z-$tube population has a negative 
kurtosis, the narrow superimposed peak from the late-formed disk gives the 
full LOSVD a net positive $h_{4}$.  This positive $h_{4}$ is greatly enhanced 
for the $z$-tube orbits alone.  In the outer dotted pixel whose LOSVD is 
shown in the middle row, the disk component is somewhat weaker and the 
negative kurtosis of the old $z$-tube population wins, giving the full 
distribution a net negative $h_{4}$.

This example demonstrates how we may get a wide spread in $h_{4}$, 
including strong positive as well as negative values, at high 
$v / \sigma$.  We caution the reader that in this situation $h_{4}$ is 
quite sensitive to small changes in the disk strength, and the GH fitting is
sensitive to noise, so care must be taken in quantitative comparisons between
simulations and observations.

In the $h_{4}-v/\sigma$ diagram the re-mergers are once again more 
concentrated around the origin, for the same reason discussed in the previous
subsection.  The smaller non-Gaussian moments of the re-mergers at fixed 
$v / \sigma$ suggest one way to separate re-merger remnants from other slow 
rotators.  The re-mergers tend to have LOSVDs with positive $h_{4}$, also an
expected signature of violent relaxation, which generally results in a
radially-biased distribution function \citep{BandT}.

Face-on projections generically give high positive $h_{4}$ values, as both the
box and $z-$tube orbits are strongly peaked around $v=0$ in this projection.
Particularly strong positive $h_{4}$ values can arise in face-on 
views of gas-rich remnants since the latest-formed stars form a narrow peak
centered on $v=0$, as illustrated in the third row of Fig. 4.  The bottom row 
of this figure shows the average locus of edge-on and face-projections in the 
$v / \sigma$ plane - face-on projections dominate the high-$h_{4}$, low 
$v / \sigma$ part of the diagram, while edge-on projections dominate the high 
$v / \sigma$ and low $h_{4}$ regions, as well as the high $v / \sigma$, high 
$h_{4}$ wings in the gas-rich remnants.

\section{Discussion and conclusions}

Mergers between spiral galaxies generically produce remnants with 
significantly non-Gaussian LOSVDs, owing to the presence of subcomponents
that retain memory of the initial conditions.  This effect is 
especially pronounced in mergers with $f_{gas} \gtrsim 20$\% where the 
dissipative component re-forms an embedded disk, producing a strong negative 
$h_{3}-v / \sigma$ correlation and a wide spread of $h_{4}$ values at high 
$v / \sigma$.  Both of these features are in agreement with observations of
elliptical galaxies \citep{b94,sauron12}.  

Upon closer examination the SAURON fast-rotator population appears to consist 
of two subgroups, one with a steep $h_{3}- v / \sigma$ correlation, and a 
second with low $h_{3}$ for all $v / \sigma$ \citep{sauron12}.  A gas-rich 
major merger scenario cannot account for the second group, which suggests 
that the two fast-rotator populations may differ in their formation mechanisms.

Dry mergers produce slowly rotating remnants with small $h_{3}$ and $h_{4}$,
as violent relaxation blurs features in the distribution function
and partially thermalizes the local velocity distribution.  This provides one 
way to distinguish dry merger remnants within the population of slowly 
rotating ellipticals.  The large spread in $h_{3}$ and $h_{4}$ among the 
SAURON slow rotators \citep{sauron12} suggests that the slow rotators 
are {\it not} all dry merger remnants, as has been proposed to explain the 
fast-slow rotator dichotomy \citep{gg06,k08}.

At fixed redshift a tight correlation is observed between the gas fraction and 
stellar mass of late-type galaxies, with $f_{gas}$ varying from $\sim$50\% at
M$_{star}=10^{10}$ M$_{\odot}$ to $\sim$10\% at M$_{star}=3 \times 10^{11}$ 
M$_{\odot}$ at $z \sim 2$ (\citealt{s09} and references therein).  The 
$f_{gas}$ labels in Figs. 1 and 3 may therefore be taken loosely as a proxy 
for galaxy mass, and our gas-rich merger model predicts a variation in the 
$h_{3,4}- v / \sigma$ relations with galaxy luminosity.

In this paper we considered only a very restricted set of merger models, 
namely binary 1:1 mergers of bulgeless disks.  The embedded disk component in 
our remnants may be artificially pronounced because real galaxies would suffer 
subsequent harassment through minor mergers, interactions with satellites, 
and secular processes.  Some of the effects of increasing $f_{gas}$ (e.g. 
higher $v / \sigma$ and $h_{3}$) can also be achieved by decreasing the 
merger mass ratio \citep{bour05,gg06,naab06,jess07,jess09}, and it is 
important to disentangle these trends.  Extensions of this work include 
varying the mass ratios and bulge fractions, simulating 
cosmologically-motivated merger sequences \citep{bur08,v09}, and including 
secular heating processes.  

Ultimately we hope to make rigorous statistical comparisons between large 
libraries of simulations and IFS data.  This will require a careful 
consideration of observational selection biases as well as systematics 
associated with the GH fitting procedure in the presence of observational
noise as well as particle discreteness in the simulations.  
This study motivates such work by illustrating how detailed 2D kinematic 
information can reveal the underlying orbit structure of elliptical galaxies, 
and place stronger constraints on models of their formation. 

\begin{acknowledgments}
We would like to thank Bart Willems for technical help, and Glenn van de Ven 
for useful discussions.  This work was supported in part by a Lindheimer 
Postdoctoral Fellowship at Northwestern University.  Computations were 
performed on the Fugu computer cluster funded by NSF MRI grant PHY-0619274 to 
Vicky Kalogera, and the Sauron computer cluster at the parallel computing 
center of the Institute for Theory and Computation at the Harvard-Smithsonian
Center for Astrophysics.
\end{acknowledgments}

\end{document}